\documentstyle[12pt]{article}
\title{Once  more  about  the $\omega \to 3\pi$  contact  term}

\author{E.~A.~Kuraev$^1$,
Z.~K.~Silagadze$^2$ \vspace*{5mm}\\
$^1$ \small\em Joint Institute for Nuclear Research, 141980 Dubna,
Moscow region, Russia \\[3mm]
$^2$ \small \em Institute of Nuclear Physics, Prospect Nauki, 11,
630090, Novosibirsk, Russia}

\date{}

\begin{document}

\maketitle

\begin{abstract}
The manifestations of the $\omega \to 3\pi$  contact  term  and its
unitary partners are investigated in the framework of the  chiral
effective lagrangian theory with vector mesons. We conclude that
nowadays  the existence and magnitude of  the  contact  term  can be
extracted neither from theory, nor experiment. The theoretical
uncertainty  is caused by the one-loop corrections. Some speculations
about them lead to the generalized KSRF relation
$\frac{f_\pi^2g^2_{\rho \pi \pi}}{m_\rho^2}=\frac{m_K}{2\sqrt{2}
\pi f_\pi}$.
\end{abstract}

\section{Introduction}

  The experimental study of the   $e^- e^+ \to 3\pi$ reaction \cite{1}
has confirmed the Gell-Mann, Sharp, Wagner suggestion \cite{2} that the
$\omega \to 3\pi$  transition is dominated by the $\omega \rho \pi$
pole diagram, though the experimental accuracy is not sufficient at
present to exclude completely the existence of the possible contact
term. This four-point contact term was discussed on quite general
grounds \cite{3,4}, inspired by dispersion theory and current-algebra.
No reason was found to neglect it, but its magnitude remained undefined
until Rudaz had remarked \cite{5} that one needs quite definite contact
term to satisfy simultaneously the KSRF relation \cite{6} and the low
energy theorem \cite{7} concerning $\pi \to 2\gamma$ and $\gamma \to
3\pi$  amplitudes.

 Meantime, Witten's topological reinterpretation \cite{8} of the
Wess-Zumino \cite{9} chiral anomaly \cite{10} stimulated a renewal of
interest in effective chiral lagrangian theories  \cite{11}. A plenty of
models were suggested ( see f.e. \cite{12} - \cite{16}), especially for
including  vector  (and  axial-vector)  degrees  of  freedom,   with
attempts \cite{17} to derive the corresponding effective
(non-renormalizable ) lagrangians  directly  from  QCD .

  Although it is commonly believed nowadays that a chiral perturbation
theory \cite{18} gives a suitable and phenomenologically successful
framework for the low energy meson physics, some specific suppositions
about vector mesons \cite{12,14} is also interesting, because they
reduce the number of phenomenological constants in the theory, so
raising its predictability.

  Namely, in \cite{12} Kaymakcalan, Rajeev and Schechter introduced
vector and axial-vector mesons as gauge bosons of local $SU(3) \times
SU(3)$ symmetry, the idea which can be traced back to Sakurai \cite{19}.
The contact term and the  $\omega \rho \pi$ coupling is fixed in their
model by demanding Bardeen's form for the chiral anomaly, but it had
been noticed soon \cite{5} that the magnitude of the contact term was
insufficient to ensure the validity of the Terentiev et al.'s low
energy theorem \cite{7}, which had been experimentally confirmed
\cite{20}.

  The situation was clarified by Brihaye, Pak and Rossi \cite{21},
who showed an elegant way how to construct counterterms \cite{22}
needed for vector mesons not to break the  low  energy  theorems of
current-algebra. Actually, in their formalism it is not obligatory
for vector mesons to be gauge bosons and throughout this paper  we
will use just such  "minimal" realization \cite{23}.

  A new stage of experiments at  Novosibirsk  VEPP-2M  storage  ring
is under way now with two modern detectors  \cite{24,25}. A few percent
accuracy is expected can be reached for many processes in the energy
range  $\sim$ 1 Gev. So a simple phenomenological vector meson dominance
picture \cite{19,24}, used earlier, becomes insufficient and it is
interesting if effective chiral lagrangian models and chiral
perturbation theory can take up this challenge.

  In this article we analyze how the model \cite{21,23} can confront
some experimental tests, with special emphasize of the effect of the
rather large $V \to 3P$ contact term. The phenomenological consequences
of current-algebra based and effective chiral lagrangian models were
thoroughly investigated  \cite{27} - \cite{29}. Therefore we omit some
technical details which can be found in the cited literature.

\section{$\Gamma (\omega \to 3\pi)$  and $e^- e^+ \to 3\pi$}

  The one of the successful predictions  of   [12]   was  a  correct
$\omega \to 3\pi$  decay width, especially compared with last
experimental results [30,31] . Adjusting  [21]  the $\omega \to 3\pi$
contact term for low energy theorem [7]  to be valid,  we  end  with
four  times  larger magnitude for it and, as a result too small
$\Gamma (\omega \to 3\pi)$ , as  will be shown below .

  Defining the $ \omega_\mu (Q)\to \pi^+(q_+) \pi^- (q_-) \pi^0 (q_0)$
amplitude as
\begin{eqnarray} \label{eq1}
   M_{\mu}&=&iF(s_{12},s_{13},s_{23})\epsilon_{\mu\nu\lambda\sigma}q_+^{\nu}
q_-^{\lambda}q_0^{\sigma}, \qquad s_{ij}=(q_i+q_j)^2,
\end{eqnarray}
a standard calculation gives the formula for the decay width
\begin{eqnarray} \label{eq2}
\Gamma(\omega\rightarrow3\pi)&=&\frac{M}{768\pi^3}\int_{x_{min}}^{x_{max}}
dx\int_{y_{min}}^{y_{max}} dy G(x,y)|M^3 F(x,y)|^2, \\ \nonumber
F(x,y) &=& F(s_{12},s_{13},s_{23}) \; ,
\end{eqnarray}
where $m=m(\pi^{\pm}),m_0=m(\pi^0),M=m(\omega),x=\frac{E_+}{M},
y=\frac{E_-}{M}$,
\begin{eqnarray}
G(x,y)=4\left (x^2-\frac{m^2}{M^2}\right )\left (y^2-\frac{m^2}{M^2}
\right )- \left (1-2x-2y+2xy+\frac{2m^2-m_0^2}{M^2}\right )^2
\label{eq3} \end{eqnarray}
and
\begin{eqnarray}
&& x_{min}=\frac{m}{M},\quad x_{max}=\frac{1}{2}\left (1-
\frac{m_0(2m+m_o)}{M^2} \right ),
\\ \nonumber &&  y_{max,min}=\frac{1}{2\left (1-2x+\frac{m^2}{M^2}
\right )} \left \{(1-x) \left (1-2x+\frac{2m^2-m_0^2}{M^2}\right ) \right .
\\ \nonumber && \qquad
\pm \left . \left [ \left (x^2-\frac{m^2}{M^2}\right ) \left (1-2x+
\frac{m_0(2m-m_0)}{M^2}\right)
\left (1-2x-\frac{m_0(2m+m_0)}{M^2}\right )\right ]^{\frac{1}{2}} \right \} .
\label{eq4} \end{eqnarray}
The expression for the invariant amplitude  F  can be obtained  from
the diagrams

%-----------------fig-----------------------------------
\vspace{.7cm}
\unitlength=1.0mm
\special{em:linewidth 0.4pt}
\linethickness{0.4pt}
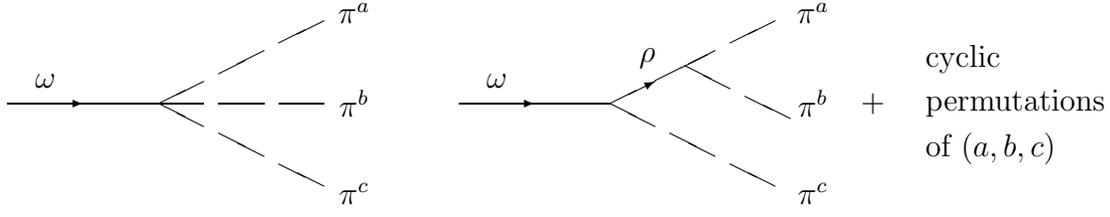
\begin{picture}(150.00,42.00)
\put(00.00,30.00){\vector(1,0){10.00}}
\put(10.00,30.00){\line(1,0){16.00}}
\put(20.00,30.00){\line(2,1){6.00}}
\put(20.00,30.00){\line(2,-1){6.00}}
\put(28.00,30.00){\line(1,0){6.00}}
\put(28.00,34.00){\line(2,1){6.00}}
\put(28.00,26.00){\line(2,-1){6.00}}
\put(36.00,30.00){\line(1,0){6.00}}
\put(36.00,38.00){\line(2,1){6.00}}
\put(36.00,22.00){\line(2,-1){6.00}}
\put(5.00,33.00){\makebox(0,0)[cc]{$\omega$}}
\put(46.00,42.00){\makebox(0,0)[cc]{$\pi^a$}}
\put(46.00,30.00){\makebox(0,0)[cc]{$\pi^b$}}
\put(46.00,18.00){\makebox(0,0)[cc]{$\pi^c$}}

\put(60.00,30.00){\vector(1,0){10.00}}
\put(70.00,30.00){\line(1,0){10.00}}

\put(80.00,30.00){\vector(2,1){6.00}}
\put(86.00,33.00){\line(2,1){8.00}}
\put(85.00,36.00){\makebox(0,0)[cc]{$\rho$}}
\put(96.00,38.00){\line(2,1){6.00}}
\put(90.00,35.00){\line(2,-1){6.00}}
\put(98.00,31.00){\line(2,-1){6.00}}

\put(65.00,33.00){\makebox(0,0)[cc]{$\omega$}}
\put(107.00,42.00){\makebox(0,0)[cc]{$\pi^a$}}
\put(107.00,30.00){\makebox(0,0)[cc]{$\pi^b$}}

\put(80.00,30.00){\line(2,-1){6.00}}
\put(88.00,26.00){\line(2,-1){6.00}}
\put(96.00,22.00){\line(2,-1){6.00}}
\put(107.00,18.00){\makebox(0,0)[cc]{$\pi^c$}}

\put(115.00,30.00){\makebox(0,0)[cc]{$+$}}
\put(122.00,36.00){\makebox(0,0)[lc]{cyclic}}
\put(122.00,30.00){\makebox(0,0)[lc]{permutations}}
\put(122.00,24.00){\makebox(0,0)[lc]{of $(a,b,c)$}}
\end{picture}

\nopagebreak
\vspace{.1cm}

\noindent
and looks like
\begin{eqnarray}
|M^3F(x,y)|^2=\left (\frac{3}{4\pi^2}\right )^2
\left (\frac{M}{f_{\pi}}\right )^6
\left (\frac{M_{\rho}}{f_{\pi}}\right )^2
\alpha_K |1-3\alpha_K-\alpha_K H(x,y)|^2,
\nonumber \end{eqnarray}
where
\begin{eqnarray}
H(x,y)=R_{\rho}(Q_0^2)+R_{\rho}(Q_+^2)+R_{\rho}(Q_-^2)
\label{eq5} \end{eqnarray}
is defined through the  $\rho$-meson  Breit-Wigner propagators
\begin{eqnarray}
R_V(Q^2)=\left [\frac{Q^2}{M_V^2}-1+i\frac{\Gamma_V(Q^2)}{M_V}\right ]^{-1}
\label{eq6} \end{eqnarray}
and
\begin{eqnarray}
Q_0^2=M^2(2x+2y-1)+m_0^2 \; , \; Q_+^2=M^2(1-2y)+m^2
\; , \; Q_-^2=M^2(1-2x)+m^2 \, .
\label{eq7} \end{eqnarray}
As  for $\alpha_K $, it is defined as
\begin{eqnarray}
\alpha_K=\left (\frac{f_{\pi}g_{\rho\pi\pi}}{M_{\rho}}\right )^2 \, ,
\label{eq8} \end{eqnarray}
$\alpha_K=\frac{1}{2}$ being  the  KSRF  relation  [6].

   For  off-mass-shell  resonance widths  we  assume  that  they  are
proportional to the main decay channel phase space. For example:
\begin{eqnarray}
\Gamma_{\rho}(Q^2)=\Gamma_{\rho} \frac{M_{\rho}^2}{Q^2}
\left (\frac{Q^2-4m^2}{M_{\rho}^2-4m^2}\right )^{\frac{3}{2}},
\label{eq9} \end{eqnarray}
\noindent
where $\Gamma_{\rho}=151 $ MeV.

   Taking for the other parameters $ f_{\pi}=93 MeV$, $m_0=m=140 MeV$,
$M=782 MeV$, $M_{\rho}=768 MeV$
and $\alpha_K= 0.55 $ (which corresponds to $\frac{g_{\rho\pi\pi}^2}
{4\pi}=3$), we get $\Gamma(\omega\rightarrow3\pi)=4.9 Mev $.
The experimental value is  [32] $\Gamma_{exp}(\omega\rightarrow3\pi)=
(7.49\pm 0.14) Mev $. If we take four times smaller contact term  from
[12] , we get almost experimental width: $\Gamma_{[12]}(\omega
\rightarrow3\pi)=7.3 Mev $ and if we drop the contact term altogether ,
the width increases  up to $ 8.4 MeV $.

  Taking into account a small deviation $\epsilon= 3.4^\circ $ [33]  from
the ideal $\omega-\phi $ mixing, the following predictions for the
$\phi\rightarrow 3\pi $  decay width can be get also :

\vspace{.5cm} \hspace{.5cm}
\begin{tabular}{|c|c|}
\hline
model           & $\Gamma (\phi \to 3\pi)$ MeV  \\ \hline
[21,23]         & 0.67                          \\ \hline
[12]            & 0.79                          \\ \hline
no contact term & 0.84                          \\ \hline
experiment [31] & 0.63 $\pm$ 0.04               \\ \hline
\end{tabular}
\vspace{.3cm}

Of course , these values  depend  on  the  details  of  the  unitary
symmetry breaking [34] ( for some new ideas about the $\omega - \phi $
mixing problem see [35] ), and so are not a clear test for the chiral
effective theories.

  Somewhat small $\Gamma(\omega\rightarrow 3\pi) $ for the  correct
(from  the  low  energy
theorem's  point  of  view  )  contact  term,  maybe  indicate   the
importance of the one-loop and radial excitations corrections. Their
magnitude can be estimated according to [36] by dual model  inspired
change
\begin{eqnarray}
R_{\rho}(x,y,Q^2)\rightarrow R_{\rho}(x,y,Q^2)
\left (\frac{F_V(Q^2)}{F_V(0)}\right )^2 \; ,
\nonumber \end{eqnarray}
where
\begin{eqnarray}
F_V(Q^2)=\Gamma(\beta-1)\frac{\Gamma(1-\alpha'(Q^2-M_{\rho}^2))}
{\Gamma(\beta-1-
\alpha'(Q^2-M_{\rho}^2))} \quad , \quad \alpha'=\frac{1}{2M_{\rho}^2}
\quad , \quad \beta \approx 2.33 \; .
\nonumber \end{eqnarray}
This increases $\Gamma(\omega\rightarrow 3\pi)$ from $ 4.9  MeV $  up
to $6.7  MeV$ in  the Brihaye, Pak, Rossi model [21,23].

  Closely  related to  the $\omega\rightarrow 3\pi $ transition  is  the
$e^+ e^-\rightarrow 3\pi$
process [37]. Its cross-section is
\begin{eqnarray}
\sigma(e^+ e^-\rightarrow 3\pi)
=\frac{\alpha}{192\pi^2 s}\int_{x_{min}}^{x_{max}}dx
\, \int_{y_{min}}^{y_{max}} dy \, G(x,y)|(2E)^3 F_{3\pi}(x,y)|^2 \; ,
\label{eq10} \end{eqnarray}
where $ s=(2E)^2$, $x=\frac{E_+}{2E}$, $y=\frac{E_-}{2E}$
and $x_{min,max}$, $y_{min,max}$, $G(x,y)$  are
given by  (3), (4)  with change $ M \rightarrow 2E, E $ being beam energy.
$ F_{3\pi}$  formfactor has the following form (here and later the
coupling constants from [23] is assumed, if not otherwise stated )
\begin{eqnarray}
&& M_\mu(\gamma\rightarrow \pi^+ \pi^- \pi^o)
= -i\epsilon_{\mu\nu\sigma\tau}q_+^{\nu}q_-^{\sigma}q_o^{\tau}
F_{3\pi}(s_{12},s_{13},s_{23}) \quad ,
\\ \nonumber &&
|(2E)^3 F_{3\pi}(x,y)|^2 =  \\ \nonumber &&
\frac{3\alpha}{4\pi^3} \left (\frac{2E}
{f_{\pi}}\right )^6 |\sin \theta \cos\epsilon R_{\omega}(s)
-\cos\theta \sin \epsilon R_{\phi}(s)|^2 |1-3\alpha_K-\alpha_K H(x,y)|^2.
\label{eq11} \end{eqnarray}
Recall $ M \rightarrow 2E $ substitution in the definition of $ H(x,y) $
from (5) and (7). $\, \theta $ is the $\omega - \phi $ mixing angle:
\begin{eqnarray}
\omega=\cos \theta \omega_{(1)}+\sin \theta \omega_{(8)} \quad , \quad
\phi= \cos \theta \omega_{(8)}-\sin \theta \omega_{(1)} \; ,
\nonumber \end{eqnarray}
and $\epsilon=\theta-\arcsin (\frac{1}{\sqrt{3}})=3.4^\circ $ÄÄ its
departure from the "ideal" mixing.

  Numerical calculations give the following resonance cross-sections
(when $  s=M_V^2 $ ) :

\begin{tabular}{|c|c|c|c|c|}
\hline
model                     &[21,23]&[12] &no contact term
& experiment                       \\ \hline
$\sigma_{\omega},\ \mu b$ &0.99   &1.47 & 1.71
&1.54 $\pm$ 0.03 $\pm$ 0.12, [39]  \\ \hline
$\sigma_{\phi},\ \mu b$ &0.58   &0.69 & 0.75
&0.66 $\pm$ 0.04, [40]             \\ \hline
\end{tabular}
\vspace{.3cm}

As expected , the situation is much the same as for $\Gamma(\omega
\rightarrow 3\pi) $ . Fig.1  shows calculated cross-sections between
$\omega $  and $\phi $   resonances ( again  [12] fits experiment better,
than  [21,23] ), and Fig.2 shows the same after $\phi $-meson.

  As we  see,  above   $\phi$-meson  chiral-model  predictions  disagrees
significantly  with  experiment.  Of  course,  the  assumption  that
resonance  widths,  being  normalized  at  their  physical   values,
increase as the main decay channel phase  space,  is  not  the  best
thing to be done when we are so far from the $\omega$-meson peak [40]. But
even  disregarding  completely  a $Q^2$--dependence  of $\Gamma_V $
in the Breit-Wigner propagators, we get only a factor about 3, not
enough to remove discrepancy.

  Maybe this experimental result can't be explained  without  radial
excitations ($\rho(1450)$ for example). In any case, we see  that  above
$1 GeV$ predictions of chiral effective  theory  must  be  dealt  with
caution. Nevertheless, below we consider some  unitary  partners  of
$ \omega \to 3\pi$ and $e^- e^+ \to 3\pi$, such as
$ K^* \to  K\pi\pi $ [27,28],
$ e^- e^+\to \pi K \bar K $ [41],
$\eta \to \pi\pi \gamma$ [27,28,42]  and
$ e^- e^+\rightarrow \pi \pi\eta $  [43].

\section{$ \Gamma(K^*\rightarrow K \pi \pi) $}

  Because of isospin and charge conjugation  invariance,  only $ K^{*+} $
decay modes can be considered. Needed formulas are the same  as  for
$ \Gamma(\omega\rightarrow 3 \pi ) $  with obvious changes $M \rightarrow
M_{K^*} , m_0\rightarrow m(K) $ . The contributing diagrams are

%-----------------fig-----------------------------------
\vspace{.7cm}
\unitlength=1.0mm
\special{em:linewidth 0.4pt}
\linethickness{0.4pt}
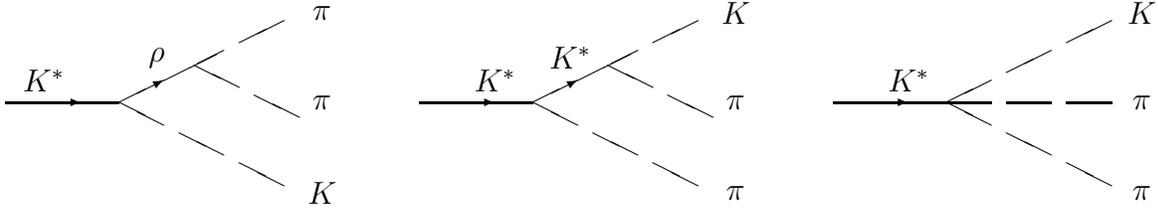
\begin{picture}(150.00,42.00)
\put(00.00,30.00){\vector(1,0){10.00}}
\put(10.00,30.00){\line(1,0){5.00}}
\put(15.00,30.00){\vector(2,1){6.00}}
\put(21.00,33.00){\line(2,1){8.00}}
\put(20.00,36.00){\makebox(0,0)[cc]{$\rho$}}
\put(31.00,38.00){\line(2,1){6.00}}
\put(25.00,35.00){\line(2,-1){6.00}}
\put(33.00,31.00){\line(2,-1){6.00}}
\put(5.00,33.00){\makebox(0,0)[cc]{$K^{*}$}}
\put(42.00,42.00){\makebox(0,0)[cc]{$\pi$}}
\put(42.00,30.00){\makebox(0,0)[cc]{$\pi$}}
\put(42.00,18.00){\makebox(0,0)[cc]{$K$}}
\put(15.00,30.00){\line(2,-1){6.00}}
\put(23.00,26.00){\line(2,-1){6.00}}
\put(31.00,22.00){\line(2,-1){6.00}}

\put(55.00,30.00){\vector(1,0){10.00}}
\put(65.00,30.00){\line(1,0){5.00}}
\put(70.00,30.00){\vector(2,1){6.00}}
\put(76.00,33.00){\line(2,1){8.00}}
\put(75.00,36.00){\makebox(0,0)[cc]{$K^{*}$}}
\put(86.00,38.00){\line(2,1){6.00}}
\put(80.00,35.00){\line(2,-1){6.00}}
\put(88.00,31.00){\line(2,-1){6.00}}
\put(65.00,33.00){\makebox(0,0)[cc]{$K^{*}$}}
\put(97.00,42.00){\makebox(0,0)[cc]{$K$}}
\put(97.00,30.00){\makebox(0,0)[cc]{$\pi$}}
\put(97.00,18.00){\makebox(0,0)[cc]{$\pi$}}
\put(70.00,30.00){\line(2,-1){6.00}}
\put(78.00,26.00){\line(2,-1){6.00}}
\put(86.00,22.00){\line(2,-1){6.00}}

\put(110.00,30.00){\vector(1,0){10.00}}
\put(120.00,30.00){\line(1,0){11.00}}
\put(125.00,30.00){\line(2,1){6.00}}
\put(125.00,30.00){\line(2,-1){6.00}}
\put(133.00,30.00){\line(1,0){6.00}}
\put(133.00,34.00){\line(2,1){6.00}}
\put(133.00,26.00){\line(2,-1){6.00}}
\put(141.00,30.00){\line(1,0){6.00}}
\put(141.00,38.00){\line(2,1){6.00}}
\put(141.00,22.00){\line(2,-1){6.00}}
\put(120.00,33.00){\makebox(0,0)[cc]{$K^{*}$}}
\put(151.00,42.00){\makebox(0,0)[cc]{$K$}}
\put(151.00,30.00){\makebox(0,0)[cc]{$\pi$}}
\put(151.00,18.00){\makebox(0,0)[cc]{$\pi$}}
\end{picture}

\nopagebreak
\vspace{.3cm}

\noindent
and  corresponding  formfactors  look  like  (for  the  model   [12]
change $1-3\alpha_K $  to $ 1-3\alpha_K+\frac{3}{2}\alpha_K^2 $  )
\begin{eqnarray}
&& F(K^*\rightarrow K^0\pi^0\pi^+)= \\ \nonumber &&
- \frac{g_{\rho\pi\pi}}{2^{\frac{3}{2}}\pi^2f_{\pi}^3}
\left [1-3\alpha_K-\frac{3}{4}\alpha_K \left (2 R_{\rho}(Q_0^2)+
\frac{m_{\rho}^2}{m_{K^*}^2}
(R_{K^*}(Q_+^2)+R_{K^*}(Q_-^2))\right )\right ], \\ \nonumber &&
F(K^*\rightarrow K^+\pi^-\pi^+)=\frac{g_{\rho\pi\pi}}{4\pi^2f_{\pi}^3}
\left [1-3\alpha_K-\frac{3}{2}\alpha_K \left ( R_{\rho}(Q_0^2)+
\frac{m_{\rho}^2}{m_{K^*}^2} R_{K^*}(Q_+^2)\right ) \right ].
\label{eq12} \end{eqnarray}
The third formfactor satisfies relation
\begin{eqnarray}
F(K^*\rightarrow K^+\pi^0\pi^0)=F(K^*\rightarrow K^+\pi^-\pi^+)+
\frac{1}{\sqrt{2}} F(K^*\rightarrow K^0\pi^0\pi^+) \; ,
\label{eq13} \end{eqnarray}
expected from the isospin invariance.
  The  numerical  results  are  collected  below  (we   have   taken
$ m_0= m(K)=500 MeV \quad , \quad M_{K^*}=890 MeV \quad , \quad
\Gamma_{K^*}=50 MeV $).

\vspace{.5cm}
\hspace{.5cm}
\begin{tabular}{|c|c|c|c|}
\hline
model                              &[21,23] &[12]    &no contact term
\\ \hline
$\Gamma (K^{*+} \to K^0\pi^0\pi^+)$&11.4 keV&17.3 keV& 20.3 keV
\\ \hline
$\Gamma (K^{*+} \to K^+\pi^+\pi^-)$& 5.7 keV& 8.7 keV& 10.2 keV
\\ \hline
$\Gamma (K^{*+} \to K^+\pi^0\pi^0)$&0.03 keV&0.03 keV& 0.03 keV
\\ \hline
\end{tabular}
\vspace{.3cm}

 Any choice of the contact  term  is  compatible  with  the  current
experimental  bound  [32]  on   the   sum   of   all   three   modes
$ \Gamma(K^*\rightarrow K\pi\pi)< 35 keV $.

\section{$ e^- e^+\rightarrow  K \bar K \pi $  near the threshold}

  This reaction was not yet observed for the energies $ s \sim
(1 GeV)^2 $ . it is interesting if the new  VEPP-2M  experiments can
see  them.  Our results show that the expected cross-sections are several
picobarns, so their investigation is not a simple, though  possible  task
for such a kind of storage ring as VEPP-2M.

  There are many diagrams contributing  in  this  process.  For  the
ideal $\omega-\phi $ mixing, they are listed below:

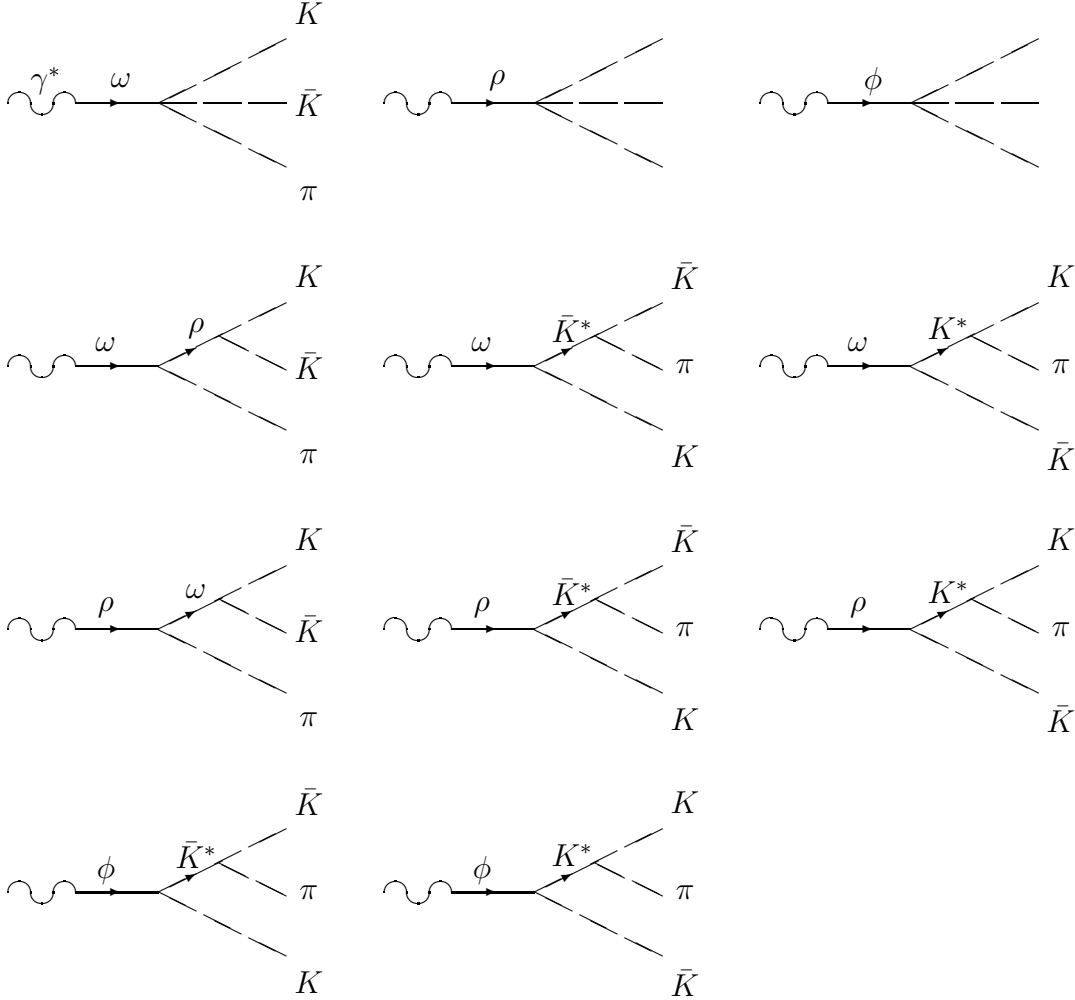
\nopagebreak
%-----------------fig-----------------------------------
\vspace{.7cm}
\unitlength=1.0mm
\special{em:linewidth 0.4pt}
\linethickness{0.4pt}
\begin{picture}(150.00,150.00)
\put(01.50,130.00){\oval(3.00,3.00)[t]}
\put(04.50,130.00){\oval(3.00,3.00)[b]}
\put(07.50,130.00){\oval(3.00,3.00)[t]}
\put(9.00,130.00){\vector(1,0){6.00}}
\put(15.00,130.00){\line(1,0){10.00}}
\put(26.00,130.00){\line(1,0){5.00}}
\put(32.00,130.00){\line(1,0){5.00}}
\put(20.00,130.00){\line(2,1){5.00}}
\put(26.00,133.00){\line(2,1){5.00}}
\put(32.00,136.00){\line(2,1){5.00}}
\put(20.00,130.00){\line(2,-1){5.00}}
\put(26.00,127.00){\line(2,-1){5.00}}
\put(32.00,124.00){\line(2,-1){5.00}}
\put(5.00,133.00){\makebox(0,0)[cc]{$\gamma^{*}$}}
\put(15.00,133.00){\makebox(0,0)[cc]{$\omega$}}
\put(40.00,142.00){\makebox(0,0)[cc]{$K$}}
\put(40.00,130.00){\makebox(0,0)[cc]{$\bar{K}$}}
\put(40.00,118.00){\makebox(0,0)[cc]{$\pi$}}

\put(51.50,130.00){\oval(3.00,3.00)[t]}
\put(54.50,130.00){\oval(3.00,3.00)[b]}
\put(57.50,130.00){\oval(3.00,3.00)[t]}
\put(59.00,130.00){\vector(1,0){6.00}}
\put(65.00,130.00){\line(1,0){10.00}}
\put(76.00,130.00){\line(1,0){5.00}}
\put(82.00,130.00){\line(1,0){5.00}}
\put(70.00,130.00){\line(2,1){5.00}}
\put(76.00,133.00){\line(2,1){5.00}}
\put(82.00,136.00){\line(2,1){5.00}}
\put(70.00,130.00){\line(2,-1){5.00}}
\put(76.00,127.00){\line(2,-1){5.00}}
\put(82.00,124.00){\line(2,-1){5.00}}
\put(65.00,133.00){\makebox(0,0)[cc]{$\rho$}}

\put(101.50,130.00){\oval(3.00,3.00)[t]}
\put(104.50,130.00){\oval(3.00,3.00)[b]}
\put(107.50,130.00){\oval(3.00,3.00)[t]}
\put(109.00,130.00){\vector(1,0){6.00}}
\put(115.00,130.00){\line(1,0){10.00}}
\put(126.00,130.00){\line(1,0){5.00}}
\put(132.00,130.00){\line(1,0){5.00}}
\put(120.00,130.00){\line(2,1){5.00}}
\put(126.00,133.00){\line(2,1){5.00}}
\put(132.00,136.00){\line(2,1){5.00}}
\put(120.00,130.00){\line(2,-1){5.00}}
\put(126.00,127.00){\line(2,-1){5.00}}
\put(132.00,124.00){\line(2,-1){5.00}}
\put(115.00,133.00){\makebox(0,0)[cc]{$\phi$}}

\put(01.50,95.00){\oval(3.00,3.00)[t]}
\put(04.50,95.00){\oval(3.00,3.00)[b]}
\put(07.50,95.00){\oval(3.00,3.00)[t]}
\put(9.00,95.00){\vector(1,0){6.00}}
\put(15.00,95.00){\line(1,0){5.00}}
\put(20.00,95.00){\vector(2,1){5.00}}
\put(25.00,97.50){\line(2,1){6.00}}
\put(32.00,101.00){\line(2,1){5.00}}
\put(20.00,95.00){\line(2,-1){5.00}}
\put(26.00,92.00){\line(2,-1){5.00}}
\put(32.00,89.00){\line(2,-1){5.00}}
\put(28.00,99.00){\line(2,-1){4.00}}
\put(33.00,96.50){\line(2,-1){4.00}}
\put(25.00,100.00){\makebox(0,0)[cc]{$\rho$}}
\put(13.00,98.00){\makebox(0,0)[cc]{$\omega$}}
\put(40.00,107.00){\makebox(0,0)[cc]{$K$}}
\put(40.00,95.00){\makebox(0,0)[cc]{$\bar{K}$}}
\put(40.00,83.00){\makebox(0,0)[cc]{$\pi$}}

\put(51.50,95.00){\oval(3.00,3.00)[t]}
\put(54.50,95.00){\oval(3.00,3.00)[b]}
\put(57.50,95.00){\oval(3.00,3.00)[t]}
\put(59.00,95.00){\vector(1,0){6.00}}
\put(65.00,95.00){\line(1,0){5.00}}
\put(70.00,95.00){\vector(2,1){5.00}}
\put(75.00,97.50){\line(2,1){6.00}}
\put(82.00,101.00){\line(2,1){5.00}}
\put(70.00,95.00){\line(2,-1){5.00}}
\put(76.00,92.00){\line(2,-1){5.00}}
\put(82.00,89.00){\line(2,-1){5.00}}
\put(78.00,99.00){\line(2,-1){4.00}}
\put(83.00,96.50){\line(2,-1){4.00}}
\put(75.00,100.00){\makebox(0,0)[cc]{$\bar{K}^{*}$}}
\put(63.00,98.00){\makebox(0,0)[cc]{$\omega$}}
\put(90.00,107.00){\makebox(0,0)[cc]{$\bar{K}$}}
\put(90.00,95.00){\makebox(0,0)[cc]{$\pi$}}
\put(90.00,83.00){\makebox(0,0)[cc]{$K$}}

\put(101.50,95.00){\oval(3.00,3.00)[t]}
\put(104.50,95.00){\oval(3.00,3.00)[b]}
\put(107.50,95.00){\oval(3.00,3.00)[t]}
\put(109.00,95.00){\vector(1,0){6.00}}
\put(115.00,95.00){\line(1,0){5.00}}
\put(120.00,95.00){\vector(2,1){5.00}}
\put(125.00,97.50){\line(2,1){6.00}}
\put(132.00,101.00){\line(2,1){5.00}}
\put(120.00,95.00){\line(2,-1){5.00}}
\put(126.00,92.00){\line(2,-1){5.00}}
\put(132.00,89.00){\line(2,-1){5.00}}
\put(128.00,99.00){\line(2,-1){4.00}}
\put(133.00,96.50){\line(2,-1){4.00}}
\put(125.00,100.00){\makebox(0,0)[cc]{$K^{*}$}}
\put(113.00,98.00){\makebox(0,0)[cc]{$\omega$}}
\put(140.00,107.00){\makebox(0,0)[cc]{$K$}}
\put(140.00,95.00){\makebox(0,0)[cc]{$\pi$}}
\put(140.00,83.00){\makebox(0,0)[cc]{$\bar{K}$}}

\put(01.50,60.00){\oval(3.00,3.00)[t]}
\put(04.50,60.00){\oval(3.00,3.00)[b]}
\put(07.50,60.00){\oval(3.00,3.00)[t]}
\put(9.00,60.00){\vector(1,0){6.00}}
\put(15.00,60.00){\line(1,0){5.00}}
\put(20.00,60.00){\vector(2,1){5.00}}
\put(25.00,62.50){\line(2,1){6.00}}
\put(32.00,66.00){\line(2,1){5.00}}
\put(20.00,60.00){\line(2,-1){5.00}}
\put(26.00,57.00){\line(2,-1){5.00}}
\put(32.00,54.00){\line(2,-1){5.00}}
\put(28.00,64.00){\line(2,-1){4.00}}
\put(33.00,61.50){\line(2,-1){4.00}}
\put(25.00,65.00){\makebox(0,0)[cc]{$\omega$}}
\put(13.00,63.00){\makebox(0,0)[cc]{$\rho$}}
\put(40.00,72.00){\makebox(0,0)[cc]{$K$}}
\put(40.00,60.00){\makebox(0,0)[cc]{$\bar{K}$}}
\put(40.00,48.00){\makebox(0,0)[cc]{$\pi$}}

\put(51.50,60.00){\oval(3.00,3.00)[t]}
\put(54.50,60.00){\oval(3.00,3.00)[b]}
\put(57.50,60.00){\oval(3.00,3.00)[t]}
\put(59.00,60.00){\vector(1,0){6.00}}
\put(65.00,60.00){\line(1,0){5.00}}
\put(70.00,60.00){\vector(2,1){5.00}}
\put(75.00,62.50){\line(2,1){6.00}}
\put(82.00,66.00){\line(2,1){5.00}}
\put(70.00,60.00){\line(2,-1){5.00}}
\put(76.00,57.00){\line(2,-1){5.00}}
\put(82.00,54.00){\line(2,-1){5.00}}
\put(78.00,64.00){\line(2,-1){4.00}}
\put(83.00,61.50){\line(2,-1){4.00}}
\put(75.00,65.00){\makebox(0,0)[cc]{$\bar{K}^{*}$}}
\put(63.00,63.00){\makebox(0,0)[cc]{$\rho$}}
\put(90.00,72.00){\makebox(0,0)[cc]{$\bar{K}$}}
\put(90.00,60.00){\makebox(0,0)[cc]{$\pi$}}
\put(90.00,48.00){\makebox(0,0)[cc]{$K$}}

\put(101.50,60.00){\oval(3.00,3.00)[t]}
\put(104.50,60.00){\oval(3.00,3.00)[b]}
\put(107.50,60.00){\oval(3.00,3.00)[t]}
\put(109.00,60.00){\vector(1,0){6.00}}
\put(115.00,60.00){\line(1,0){5.00}}
\put(120.00,60.00){\vector(2,1){5.00}}
\put(125.00,62.50){\line(2,1){6.00}}
\put(132.00,66.00){\line(2,1){5.00}}
\put(120.00,60.00){\line(2,-1){5.00}}
\put(126.00,57.00){\line(2,-1){5.00}}
\put(132.00,54.00){\line(2,-1){5.00}}
\put(128.00,64.00){\line(2,-1){4.00}}
\put(133.00,61.50){\line(2,-1){4.00}}
\put(125.00,65.00){\makebox(0,0)[cc]{$K^{*}$}}
\put(113.00,63.00){\makebox(0,0)[cc]{$\rho$}}
\put(140.00,72.00){\makebox(0,0)[cc]{$K$}}
\put(140.00,60.00){\makebox(0,0)[cc]{$\pi$}}
\put(140.00,48.00){\makebox(0,0)[cc]{$\bar{K}$}}

\put(01.50,25.00){\oval(3.00,3.00)[t]}
\put(04.50,25.00){\oval(3.00,3.00)[b]}
\put(07.50,25.00){\oval(3.00,3.00)[t]}
\put(09.00,25.00){\vector(1,0){6.00}}
\put(15.00,25.00){\line(1,0){5.00}}
\put(20.00,25.00){\vector(2,1){5.00}}
\put(25.00,27.50){\line(2,1){6.00}}
\put(32.00,31.00){\line(2,1){5.00}}
\put(20.00,25.00){\line(2,-1){5.00}}
\put(26.00,22.00){\line(2,-1){5.00}}
\put(32.00,19.00){\line(2,-1){5.00}}
\put(28.00,29.00){\line(2,-1){4.00}}
\put(33.00,26.50){\line(2,-1){4.00}}
\put(25.00,30.00){\makebox(0,0)[cc]{$\bar{K}^{*}$}}
\put(13.00,28.00){\makebox(0,0)[cc]{$\phi$}}
\put(40.00,37.00){\makebox(0,0)[cc]{$\bar{K}$}}
\put(40.00,25.00){\makebox(0,0)[cc]{$\pi$}}
\put(40.00,13.00){\makebox(0,0)[cc]{$K$}}

\put(51.50,25.00){\oval(3.00,3.00)[t]}
\put(54.50,25.00){\oval(3.00,3.00)[b]}
\put(57.50,25.00){\oval(3.00,3.00)[t]}
\put(59.00,25.00){\vector(1,0){6.00}}
\put(65.00,25.00){\line(1,0){5.00}}
\put(70.00,25.00){\vector(2,1){5.00}}
\put(75.00,27.50){\line(2,1){6.00}}
\put(82.00,31.00){\line(2,1){5.00}}
\put(70.00,25.00){\line(2,-1){5.00}}
\put(76.00,22.00){\line(2,-1){5.00}}
\put(82.00,19.00){\line(2,-1){5.00}}
\put(78.00,29.00){\line(2,-1){4.00}}
\put(83.00,26.50){\line(2,-1){4.00}}
\put(75.00,30.00){\makebox(0,0)[cc]{$K^{*}$}}
\put(63.00,28.00){\makebox(0,0)[cc]{$\phi$}}
\put(90.00,37.00){\makebox(0,0)[cc]{$K$}}
\put(90.00,25.00){\makebox(0,0)[cc]{$\pi$}}
\put(90.00,13.00){\makebox(0,0)[cc]{$\bar{K}$}}
\end{picture}

\nopagebreak
\vspace{.1cm}
\noindent
The formfactors can be easily derived  from  them  using  coupling
constants of [23] and have the following form:
\begin{eqnarray}
&& F(e^+e^-\rightarrow K^+ K^- \pi^0)=\frac{e}{12\pi^2 f_{\pi}^3}
[F_{\omega}+3F_{\rho}-F_{\phi}] \quad , \\ \nonumber &&
F(e^+e^-\rightarrow K^0 \bar K^0\pi^0)=\frac{e}{12\pi^2 f_{\pi}^3}
[-F_{\omega}+3F_{\rho}+F_{\phi}] \quad , \\ \nonumber &&
F(e^+e^-\rightarrow K^+ \bar K^0\pi^-)=\frac{e\sqrt{2}}{12\pi^2 f_{\pi}^3}
[F_{\omega}+3\tilde F_{\rho}-F_{\phi}] \quad ,
\label{eq14} \end{eqnarray}
\noindent
where
\begin{eqnarray} &&
F_{\omega}=R_{\omega}(s)\left [1-3\alpha_K-\frac{3}{2}\alpha_KR_{\rho}
(Q_0^2)-
\frac{3}{4} \alpha_K \frac{m_{\rho}^2}{m_{K^*}^2}(R_{K^*}(Q_+^2)+
R_{K^*}(Q_-^2))\right ] \quad , \nonumber \\ &&
F_{\rho}=R_{\rho}(s)\left [1-3\alpha_K-\frac{3}{2}\alpha_K \frac{m_
{\rho}^2}{M^2}
R_{\omega}(Q_0^2)-\frac{3}{4} \alpha_K \frac{m_{\rho}^2}{m_{K^*}^2}
(R_{K^*}(Q_+^2)+R_{K^*}(Q_-^2)) \right ] \quad , \nonumber \\ &&
F_{\phi}=R_{\phi}(s)\left [1-3\alpha_K-\frac{3}{2}\alpha_K \frac{m_
{\rho}^2}{M_{K^*}^2}
[R_{K^*}(Q_+^2)+R_{K^*}(Q_-^2)\right ] \quad ,
\nonumber \end{eqnarray}
\noindent
and
\begin{eqnarray}
\tilde F_{\rho}=\frac{3}{4} \alpha_K \frac{m_{\rho}^2}{m_{K^*}^2}R_{\rho}(s)
[R_{K^*}(Q_+^2)-R_{K^*}(Q_-^2)]
\nonumber \end{eqnarray}
$ Q_0^2,Q_+^2 $ and $ Q_-^2 $ are given by (7) with changes
\begin{eqnarray}
M^2\rightarrow s=(2E)^2 , \; m_0\rightarrow m(K), \; m\rightarrow m(K).
\nonumber \end{eqnarray}
 The formula for the cross-section is actually the same as  for  the
$ e^+ e^-\rightarrow 3\pi $ and  Fig.3-5  present the numerical results.

\section{$\Gamma(\eta\rightarrow \pi\pi\gamma) $ and $ e^+ e^-\rightarrow
\eta\pi\pi $ near the threshold}
$\eta \rightarrow \pi\pi\gamma $ decay in our models goes through the
diagrams

%-----------------fig-----------------------------------
\vspace{.7cm}
\unitlength=1.0mm
\special{em:linewidth 0.4pt}
\linethickness{0.4pt}
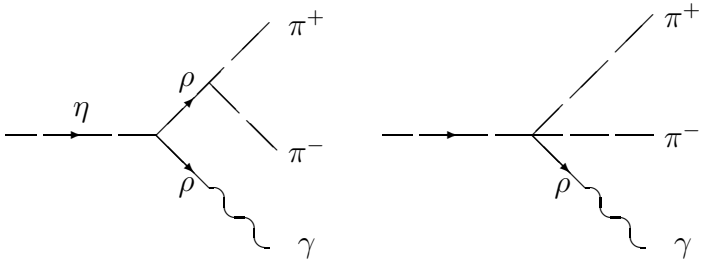
\begin{picture}(130.00,42.00)
\put(00.00,25.00){\line(1,0){4.00}}
\put(05.00,25.00){\vector(1,0){5.00}}
\put(10.00,25.00){\line(1,0){4.00}}
\put(15.00,25.00){\line(1,0){5.00}}
\put(20.00,25.00){\vector(1,1){5.00}}
\put(25.00,30.00){\line(1,1){4.00}}
\put(30.00,35.00){\line(1,1){5.00}}
\put(27.00,32.00){\line(1,-1){4.00}}
\put(32.00,27.00){\line(1,-1){4.00}}
\put(20.00,25.00){\vector(1,-1){5.00}}
\put(20.00,25.00){\line(1,-1){7.00}}
\put(27.00,16.00){\oval(4.00,4.00)[rt]}
\put(31.00,16.00){\oval(4.00,4.00)[lb]}
\put(31.00,12.00){\oval(4.00,4.00)[rt]}
\put(35.00,12.00){\oval(4.00,4.00)[lb]}
\put(10.00,28.00){\makebox(0,0)[cc]{$\eta$}}
\put(24.00,32.00){\makebox(0,0)[cc]{$\rho$}}
\put(24.00,18.00){\makebox(0,0)[cc]{$\rho$}}
\put(40.00,40.00){\makebox(0,0)[cc]{$\pi^+$}}
\put(40.00,23.00){\makebox(0,0)[cc]{$\pi^-$}}
\put(40.00,10.00){\makebox(0,0)[cc]{$\gamma$}}
\put(50.00,25.00){\line(1,0){4.00}}
\put(55.00,25.00){\vector(1,0){5.00}}
\put(60.00,25.00){\line(1,0){4.00}}
\put(65.00,25.00){\line(1,0){9.00}}
\put(75.00,25.00){\line(1,0){5.00}}
\put(81.00,25.00){\line(1,0){5.00}}
\put(70.00,25.00){\line(1,1){4.00}}
\put(75.00,30.00){\line(1,1){5.00}}
\put(81.00,36.00){\line(1,1){5.00}}
\put(70.00,25.00){\vector(1,-1){5.00}}
\put(70.00,25.00){\line(1,-1){7.00}}
\put(77.00,16.00){\oval(4.00,4.00)[rt]}
\put(81.00,16.00){\oval(4.00,4.00)[lb]}
\put(81.00,12.00){\oval(4.00,4.00)[rt]}
\put(85.00,12.00){\oval(4.00,4.00)[lb]}
\put(74.00,18.00){\makebox(0,0)[cc]{$\rho$}}
\put(90.00,41.00){\makebox(0,0)[cc]{$\pi^+$}}
\put(90.00,25.00){\makebox(0,0)[cc]{$\pi^-$}}
\put(90.00,10.00){\makebox(0,0)[cc]{$\gamma$}}
\end{picture}

\nopagebreak
\vspace{.1cm}
\noindent
and so can be considered as one more unitary partner of
$ \omega \to 3\pi $.  But here the situation is complicated
by the fact that $\eta-\eta'$ mixing  can
effect significantly the decay width.
  If we include $ \eta' $   -meson  by the nonet symmetry  prescription
[44] $ \Phi \rightarrow \Phi+\frac{1}{\sqrt{3}}\eta_{(1)} $  with
\begin{eqnarray}
\eta=\cos \theta \eta_{(8)}-\sin \theta \eta_{(1)} \quad , \quad
\eta'=\cos \theta \eta_{(1)}+\sin \theta \eta_{(8)},
\nonumber \end{eqnarray}
then get for the $ \eta \rightarrow \pi\pi\gamma $  invariant amplitude
\begin{eqnarray}
F_{\pi\pi\gamma}=\frac{e}{4\sqrt{3}\pi^2f_{\pi}^3}[\cos \theta-\sqrt{2}
\sin \theta] [1-3\alpha_K-3\alpha_KR_{\rho}(Q_0^2)].
\label{eq15} \end{eqnarray}
 Assuming $ M\rightarrow m(\eta),m=m(\pi),m_0=0 $ in (2),(3),(4) and
$ \theta=-20^\circ $  [45],
we can calculate the decay width and the results are :

\hspace{.5cm}
\begin{tabular}{|c|c|c|c|}
\hline
model                              &[21,23] &[12]  &no contact term \\
\hline
$\Gamma (\eta \to \pi \pi \gamma)$ & 93 eV  &152 eV& 183 eV         \\
\hline
\end{tabular}
\vspace{.3cm}

This   evidently   overestimates   the   experimental   width   [32]
$\Gamma_{exp}=(57\pm7) eV $. But remember that $\cos \theta-\sqrt{2}\sin
\theta \approx \sqrt{2} $
factor in  (15)
is due to the $\eta-\eta'$ mixing. So it is not clear if this discrepancy
indicates the important one-loop corrections [46]  or  more  refined
$\eta-\eta'$  mixing scheme [47] .

  $ e^+ e^-\rightarrow \eta\pi\pi $  reaction can be considered in the
same  way  as $ e^+ e^-\rightarrow 3\pi $.
 For the above  mentioned $\eta-\eta'$ mixing, the formfactor looks like
\begin{eqnarray}
F(e^+ e^-\rightarrow \eta\pi\pi)=\frac{e}{4\sqrt{3}\pi^2f_{\pi}^3}[\cos
\theta-
\sqrt{2}\sin \theta]R_{\rho}(s)[1-3\alpha_K-3\alpha_K R_{\rho}(Q_0^2)],
\nonumber \end{eqnarray}
\noindent
and corresponds to the diagrams

%-----------------fig-----------------------------------
\vspace{.7cm}
\unitlength=1.0mm
\special{em:linewidth 0.4pt}
\linethickness{0.4pt}
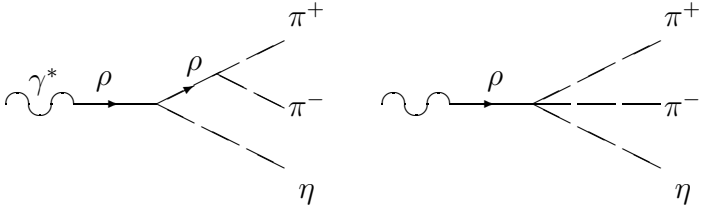
\begin{picture}(150.00,50.00)
\put(01.50,25.00){\oval(3.00,3.00)[t]}
\put(04.50,25.00){\oval(3.00,3.00)[b]}
\put(07.50,25.00){\oval(3.00,3.00)[t]}
\put(09.00,25.00){\vector(1,0){6.00}}
\put(15.00,25.00){\line(1,0){5.00}}
\put(20.00,25.00){\vector(2,1){5.00}}
\put(25.00,27.50){\line(2,1){6.00}}
\put(32.00,31.00){\line(2,1){5.00}}
\put(20.00,25.00){\line(2,-1){5.00}}
\put(26.00,22.00){\line(2,-1){5.00}}
\put(32.00,19.00){\line(2,-1){5.00}}
\put(28.00,29.00){\line(2,-1){4.00}}
\put(33.00,26.50){\line(2,-1){4.00}}
\put(05.00,28.00){\makebox(0,0)[cc]{$\gamma^{*}$}}
\put(25.00,30.00){\makebox(0,0)[cc]{$\rho$}}
\put(13.00,28.00){\makebox(0,0)[cc]{$\rho$}}
\put(40.00,37.00){\makebox(0,0)[cc]{$\pi^+$}}
\put(40.00,25.00){\makebox(0,0)[cc]{$\pi^-$}}
\put(40.00,13.00){\makebox(0,0)[cc]{$\eta$}}

\put(51.50,25.00){\oval(3.00,3.00)[t]}
\put(54.50,25.00){\oval(3.00,3.00)[b]}
\put(57.50,25.00){\oval(3.00,3.00)[t]}
\put(59.00,25.00){\vector(1,0){6.00}}
\put(65.00,25.00){\line(1,0){10.00}}
\put(76.00,25.00){\line(1,0){5.00}}
\put(82.00,25.00){\line(1,0){5.00}}
\put(70.00,25.00){\line(2,1){5.00}}
\put(76.00,28.00){\line(2,1){5.00}}
\put(82.00,31.00){\line(2,1){5.00}}
\put(70.00,25.00){\line(2,-1){5.00}}
\put(76.00,22.00){\line(2,-1){5.00}}
\put(82.00,19.00){\line(2,-1){5.00}}
\put(65.00,28.00){\makebox(0,0)[cc]{$\rho$}}
\put(90.00,37.00){\makebox(0,0)[cc]{$\pi^+$}}
\put(90.00,25.00){\makebox(0,0)[cc]{$\pi^-$}}
\put(90.00,13.00){\makebox(0,0)[cc]{$\eta$}}
\end{picture}

\nopagebreak
\vspace{.1cm}
\noindent

 The predicted cross-sections are (in picobarns) :

\hspace{.5cm}
\begin{tabular}{|c|c|c|c|c|}
\hline
\multicolumn{1}{|c|}{$2E$, GeV}
& \multicolumn{3}{c}{model} &
\multicolumn{1}{|c|}{experiment} \\ \cline{2-4}
  &  [21,23] & [12] & no contact term &  [31] \\ \hline
1.075     & 8       & 12   & 15                   & 0 $\pm$ 500     \\
\hline
1.15      & 18      & 28   & 32                   & 0 $\pm$ 500     \\
\hline
1.25      & 57      & 75   & 81                   & 200 $\pm$ 400   \\
\hline
1.325     & 133     & 162  & 176                  & 300 $\pm$ 500   \\
\hline
\end{tabular}
\vspace{.3cm}

 This process was considered earlier in [43] with similar  results.
It is premature to compare them to the  existing  experimental  data
because of a very big statistical errors.

  For higher energies it  is  known  [48]  that  the  reaction  goes
through the $\rho $ -meson radial excitations, so we don't expect that the
predictions of our chiral effective lagrangians can be  trusted  far
from the threshold.

\section{ Beyond the trees}

  As we have seen, phenomenological consequences of chiral effective
theory with correct $\omega\rightarrow 3\pi $ contact term can be
hardly considered as
successful.  This  naturally  raises  a  question   about   one-loop
corrections [49].

  The full investigation of  the  one-loop  renormalization  in  the
model [23] is out of the scope of this article. Here we  only  like
to mention that an advantage of [21]-type models, compared to [12] ,
in ability to reproduce the low  energy  theorems,  becomes  not  so
obvious when we go beyond the tree level. In  particularly,  let  us
note the curious observation that the purely pseudoscalar loops  can
restore in [12] the validity of the Terentiev et  al.'s  low  energy
theorem $ F_{3\pi}=\frac{F_{\pi}}{ef_{\pi}^2} $ [7].

  The one-loop renormalization of the $\pi\rightarrow 2\gamma $  amplitude
has  been
already considered [50]. It was found that $ F_{\pi} $  ( i.e. the
$\omega \rho \pi$ vertex)
remains unrenormalized,  as  was  expected  from  the  Adler-Bardeen
theorem about the non-renormalizability of the chiral anomaly [51].

  The contributions from the  pion-loop contained  diagrams  in  the
low energy $\gamma\rightarrow 3\pi $ amplitude are  proportional  to
$ m_{\pi}^2 $ and  can  be
neglected in the spirit  of  current-algebra.  So  it  remains  only
the $ K \bar K\rightarrow 3\pi $  Wess-Zumino anomaly contribution :

%-----------------fig-----------------------------------
\vspace{.7cm}
\unitlength=1.0mm
\special{em:linewidth 0.4pt}
\linethickness{0.4pt}
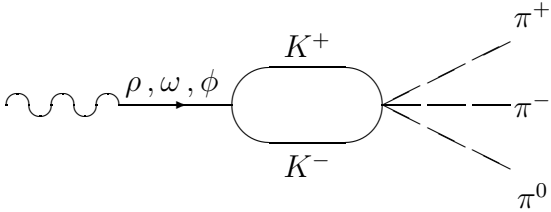
\begin{picture}(150.00,50.00)
\put(11.50,25.00){\oval(3.00,3.00)[t]}
\put(14.50,25.00){\oval(3.00,3.00)[b]}
\put(17.50,25.00){\oval(3.00,3.00)[t]}
\put(20.50,25.00){\oval(3.00,3.00)[b]}
\put(23.50,25.00){\oval(3.00,3.00)[t]}
\put(25.00,25.00){\vector(1,0){9.00}}
\put(34.00,25.00){\line(1,0){6.00}}
\put(50.00,25.00){\oval(20.00,10.00)}
\put(32.00,28.00){\makebox(0,0)[cc]{$\rho\, , \omega\, , \phi$}}
\put(50.00,33.00){\makebox(0,0)[cc]{$K^+$}}
\put(50.00,17.00){\makebox(0,0)[cc]{$K^-$}}

\put(60.00,25.00){\line(1,0){5.00}}
\put(66.00,25.00){\line(1,0){5.00}}
\put(72.00,25.00){\line(1,0){5.00}}
\put(60.00,25.00){\line(2,1){5.00}}
\put(66.00,28.00){\line(2,1){5.00}}
\put(72.00,31.00){\line(2,1){5.00}}
\put(60.00,25.00){\line(2,-1){5.00}}
\put(66.00,22.00){\line(2,-1){5.00}}
\put(72.00,19.00){\line(2,-1){5.00}}
\put(80.00,37.00){\makebox(0,0)[cc]{$\pi^+$}}
\put(80.00,25.00){\makebox(0,0)[cc]{$\pi^-$}}
\put(80.00,13.00){\makebox(0,0)[cc]{$\pi^0$}}
\end{picture}

\nopagebreak
\vspace{.1cm}
\noindent

  Using dimensional regularization, we get the following  expression
for it (assuming $ \frac{1}{\varepsilon}$ pole is absorbed by suitable
counterterm)
\begin{eqnarray}
F_{3\pi}=-\frac{3em_K^2}{4(2\pi)^4 f_{\pi}^5}\left [\ln(\frac{\pi m_K^2}
{\mu^2}) +\gamma \right ] \; ,
\label{eq16} \end{eqnarray}
$\gamma=0.5772 $ being the Euler constant.

  Together with the tree level contributions in $ F_{\pi} $  and
$ F_{3\pi} $  [5] , we get that the low energy theorem [7] is satisfied,
if
\begin{eqnarray}
-\frac{3 m_K^2}{16\pi^2f_{\pi}^2}\left [\ln{\frac{\pi m_K^2}{\mu^2}}
+\gamma \right ]+
3\alpha_K+1-3\alpha_K+\frac{3}{2}\alpha_K^2=1.
\label{eq17} \end{eqnarray}
For the renormalization scale the natural choice is [52] $\mu=M_{\rho} $.
So we get from (17) a generalized  KSRF relation
\begin{eqnarray}
\alpha_K=\frac{m_K}{2\sqrt{2}\pi f_{\pi}}[\ln(\frac{\pi m_K^2}{M_\rho^2})+
\gamma]^{\frac{1}{2}}.
\label{eq18} \end{eqnarray}
For $ m_K =494 MeV , M_{\rho} =768 MeV $ and $ f_{\pi} =93 MeV $ the  r.h.s.
gives just the experimental value  0.55 .

  Using $ SU(6) $ motivated relation [53] $\frac{g_{\rho\pi\pi}^2}{4\pi}=
\frac{2}{3}\sqrt{2}\pi \; $, one  more
interesting formula can be obtained from (18) :
\begin{eqnarray}
f_{\pi}=\frac{M_{\rho}}{4\pi}\left (\frac{6m_K}{M_{\rho}}\right )
^{\frac{1}{3}}
\left [\ln{\frac{\pi m_K^2}{M_\rho^2}}+\gamma \right ]^{\frac{1}{6}}
\approx \frac{M_{\rho}}{4\pi}\left (\frac{6m_K}{M_{\rho}}\right )
^{\frac{1}{3}}   \; .
\label{eq19} \end{eqnarray}

  We don't  know, if a very good accuracy by which (18) and (19) are
fulfilled is a mere accident, or the consistency of the vector meson
dominance, chiral loops and low energy theorems really requires such
 a kind of relations.

\section{Conclusions}

  It seems to us that  at  present  neither  theory  nor  experiment
points out to the existence and  magnitude  of  the $\omega\rightarrow
3\pi $ contact term. A relative size of the contact interactions, compared
to  the $\rho$-meson pole one, can be in principle  extracted  from  a
$ e^+ e^-\rightarrow 3\pi $
data investigating a $ \frac{d\sigma}{dx dy} $ distribution, where
$ x=\frac{E_+}{2E}$ and $  y=\frac{E_-}{2E} $
are the charge pion energy fractions :
\begin{eqnarray}
\frac{d\sigma}{dx \, dy} \sim | h_{cont.}-\alpha_K H(x,y)|^2G(x,y).
\nonumber \end{eqnarray}
For example, $ h_{cont.}=1-3\alpha_K=-\frac{1}{2} $ for  [21]   and
$ h_{cont.}
=1-3\alpha_K+\frac{3}{2}\alpha_K^2=-\frac{1}{8} $ for [12] .

  The similar analysis was already performed  [1]  with  the  result
that the Gell-Mann, Sharp, Wagner  mechanism  gives $ (85\pm15)\% $ of the
total cross-section near  the $\phi $-meson  [31],  indicating  that  the
contact term, if present, is small.

  As for the  theory, the complete one-loop analysis of  the  chiral
effective models with vector mesons is greatly  desired,  especially
in context of the vector meson dominance and low energy theorems.
 $\Gamma(\omega\rightarrow 3\pi) $  and $\sigma(e^+e^-\rightarrow 3\pi) $
are the only clear touchstones for the
$\omega\rightarrow 3\pi $ contact term, because their unitary partners,
discussed above, unfortunately suffer from  the  ambiguities  associated
with the  symmetry  breaking  and  particle  mixing  details  and  radial
excitations.

\end{document}